# Catalogue of star positions and B-magnitudes in 60th declination zone based on UkrVO Joint Digital Archive

V. N. Andruk[1], L. K. Pakuliak[1], V. V. Golovnya[1], G. O. Ivanov[1],
O.M. Yizhakevych[1], Yu.I. Protsyuk[2], S.V. Shatokhina[1]

[1] Main Astronomical Observatory of NAS of Ukraine
Acad. Zabolotnoho str., 27, Kyiv, 03680 Ukraine.
E-mail: andruk@mao.kiev.ua
[2] RI "Nikolaev Astronomical Observatory" of Ministry of Education and Science of Ukraine
Observatorna str. 1, Mykolaiv, 54030 Ukraine

**Abstract.** Catalogue of star positions and B-magnitudes based on UkrVO Joint Digital Archive has been created in $60^o$ declination zone of FONAK(FON) observational program as the first attempt to use the commercial scanners for astrometric purposes. The height of zone is $8^o$, number of involved plates is 120. Digital images of plates were obtained using Microtek ScanMaker 9800XL TMA commercial scanner with the plate resolution 1200 dpi, linear dimensions 13,000x13,000 px for plates 30x30cm. The catalogue includes 1, 263, 932 stars and galaxies down to $B \leq 16.5^m$ at the epoch 1984.76±0.50. Positions of objects are in TYCHO-2 reference frame, B-magnitudes in the system of photoelectric standards. The internal accuracy for all objects is $\sigma_{RA,DEC} = \pm 0.26"$ and $\sigma_B = \pm 0.17^m$, except stars in the $B = 8^m$ -$13^m$ interval having the errors of $\sigma_{RA,DEC} = \pm 0.13"$ and $\sigma_B = \pm 0.11^m$. The convergence of star positions with TYCHO-2 system is $\sigma_{RA,DEC} = \pm 0.06"$ (based on 93, 925 stars), the convergence with photoelectric standards system is $\sigma_B = \pm 0.16^m$ (based on 4, 458 stars). External comparison with UCAC-4 gives $\sigma_{RA,DEC} = \pm 0.34"$ (based on 1, 099, 005 cross-identified objects).

## Introduction.

The pilot version of the complete catalogue of objects in the 60th declination zone of the FON (Northern Sky Photographic Survey from 0° to 90°) project was obtained by authors several years ago. It became the first full-scale experience to use the digital material of commercial scanners for the precise positional determinations in photographic astrometry [9]. As a rule, the major part of investigations for that moment was based on the digital data from scanners, designed and built for specific photometric projects, which consequently were focused on the accuracy of photometric results. But photographic archives, collected in observatories during decades, are not only of photometric value, but they could be extremely useful in positional determinations. The expediency of this is determined at least by the fact that with few exceptions measuring technology of previous years was not possible to obtain data for all detected celestial objects on photographic plates. The photographic material was used to address highly specialized tasks on a limited range of observation. Therefore, re-use of observation archives on the current stage of development of information technology made it possible to realize their full potential.

The problems solved by the pilot version of the catalogue included as studies of commercial scanners on the suitability for accurate positional determinations and selection of the best methods of scanning and processing of digital images. As a result, the technique of combined processing of two digital images was selected as the basic one. Images are turned one to another by 90º in order to eliminate mechanical errors of the scanner. It is obvious that that technique enlarge twice the initial data amount and the processing time also [5,9].

To the end of 2014 more than 2400 digital images of FON plates were obtained in the framework of UkrVO national project [4,7,19]. One-third of this quantity was made with Microtek ScanMaker 9800XL TMA scanner, the rest – with Epson Expression 10000XL one [1,2,5,15,16,17]. All the images were processed in LINUX/MIDAS/ROMAFOT environment, specially upgraded for tasks of large field image processing and pixel coordinates and photometric estimations for all detected objects were derived [10,11]. The basic software, designed in the MAO NASU Department of Astrometry for astrometric and photometric solutions was successfully applied in practice in the series of projects [6,8,9,14,18,20].

The process of digitizing the UkrVO archives [4] was not limited to photographic surveys relatively homogeneous in quality. Plates, obtained in a variety of observational programs and received on different instruments with different methods, different structures of object images, digitized with different models of scanners were taken into processing. The variety of digitized material required constant upgrading of software and finding new approaches to its solution, including the search for means to reduce the labor content and time in mass work. Efforts resulted in developing a method of accounting scanner mechanical errors without using the 90º-turned image. The second version of the catalog of 60th zone was obtained on half the amount of negatives using this technique.

Just as for the FON program, for the 60th declination zone the principle of four-fold overlap implemented, when the centers of plates offset from each other by about four degrees in right ascension. Plates are exposed on MAO NASU DWAA (Double Wide-angle Astrograph (D/F=40/200, 103"/mm)). Linear dimensions of the most plates are 30x30 cm (8ºx8º). All the plates were scanned on Microtek ScanMaker 9800XL TMA with 1200 dpi resolution. The dimensions of digital images are up to 13, 000 x 13, 000 px (1 px=2.17").

In all the procedures the star catalogue TYCHO-2 was taken as reference. The accuracy of it is $\sigma_{RA,DEC} = \pm 0.060"$, $\sigma_\mu = \pm 0.0025"/yr$, $\sigma_m = \pm 0.10^m$ for all its 2, 539, 913 stars.

The second version of the catalogue is obtained from the processing of single scans without turning the plate by 90°. This lets to save resources for storage and processing data in half without sacrificing the accuracy of the results. Principles and stages of digitized astronegatives' processing for 60-degree zone, which are described in this paper, are now extended to the processing of Kiev part of the FON program in total.

**Preliminary studies and selection of the optimal options**

**Accuracy behavior of commercial scanners**. For digitizing of MAO photographic archive two models of scanners have been used: Microtek and Epson. Both were tested for positional and photometric accuracy. One of the plates of FON collection, having ideal visual image characteristics, was taken as a test plate. The results of its processing are given in Tables 1,2.

Table 1 shows positional errors $\sigma_{RA}$, $\sigma_{DEC}$, derived in a final reduction procedure with TYCHO-2 star catalogue as reference. Here $\sigma_{Bph}$ are photometric errors in the system of photoelectric standards $B_{pe}$ ($\sigma_{Bph}$, $n_{pe}$=644 – number of comparison stars) [12,13]. $N_{RA,DEC}$ – number of TYCHO-2 reference stars. From the Table 1 it is clear that positional and photometric errors of Epson are less than those of Mikrotek by 25 %.

The new comparison with the system of reference stars was made for averaged coordinates and magnitudes from two arrays of measurements. Its results are given in Table 2. Formally the errors of the averaged data are less by 50% and 10% than expected ones ( without taking into account the correlation between two data arrays and with it, meaning that the reference catalogue in both cases is TYCHO-2 ). The internal convergence errors for magnitudes are several times less than expected ones. This means that software developed and used for the processing permits to make the astrometric and photometric solution with a high accuracy.

Table 1. Positional errors $\sigma_{RA}$, $\sigma_{DEC}$ and magnitudes $\sigma_{Bph}$ for Epson and Microtek scanners

| $B_{ph}$ | $\sigma_{RA}$ | $\sigma_{DEC}$ | $k_{RA,DEC}$ | $\sigma_{Bph}$ | $n_{pe}$ |
|---|---|---|---|---|---|
| Epson | ± 0.087" | ±0.091" | 7,322 | ± 0.115$^m$ | 644 |
| Microtek | 0.122 | 0.114 | 7,354 | 0.141 | 644 |

**Expected accuracy, derived from two catalogues comparison**

| | | | | | |
|---|---|---|---|---|---|
| Without corr. | 0.150 | 0.146 | | 0.182 | |
| With corr. | 0.109 | 0.105 | | 0.130 | |

Table 2. An accuracy of the averaged positions and magnitudes. Distribution of errors in intervals of magnitudes

| $B_{ph}$ | $\sigma_{RA}$ | $\sigma_{DEC}$ | $\sigma_{Bph}$ | n |
|---|---|---|---|---|
| 6.58$^m$ | ± 0.042" | ±0.064" | ± 0.062$^m$ | 4 |
| 7.67 | 0.213 | 0.107 | 0.143 | 23 |
| 8.60 | 0.147 | 0.099 | 0.061 | 98 |
| 9.58 | 0.095 | 0.096 | 0.049 | 298 |
| 10.56 | 0.100 | 0.092 | 0.042 | 727 |
| 11.59 | 0.078 | 0.079 | 0.041 | 2,078 |
| 12.47 | 0.090 | 0.089 | 0.052 | 2,835 |
| 13.35 | 0.102 | 0.101 | 0.055 | 1,049 |
| 14.28 | 0.101 | 0.071 | 0.053 | 75 |
| 15.05 | 0.020 | 0.060 | 0.094 | 2 |
| 11.98 | 0.091 | 0.089 | 0.049 | 7,189 |

**Results of plate scanning in two positions.** Under the normal (or full) scanning we mean the scanning, when the Y-axis of the image ( the movement direction of the measuring CCD- line) coincides with the DEC-axis of the equatorial coordinate system. Scanning the plate at 90° turn on the scanner table suggests that the Y-axis of the image is parallel to the RA-axis.

Fig. 1 shows the results of astrometric solution for the test plate. The trend of systematic differences between obtained and catalogue positions $\Delta\alpha$, $\Delta\delta$ in relation to X,Y-axes of image is shown on the left side of the figure. Upper (a,b) and lower (c,d) panels of the figure concern the normal and turned plate positions on the table of the scanner. The distribution of the residual random differences over the plate field after the correction for scanner systematic errors and telescope aberrations is shown on the right side of Fig. 1.

The normal scanning of the test plate gives the errors of astrometric solution less by 25% than those of at 90° turn. This result is expected of the reasons that the scanner matrix movement irregularity along the Y-axis in fact causes the irregular changes of the plate image scale along the RA-axis in relation to upper and lower sides of image.

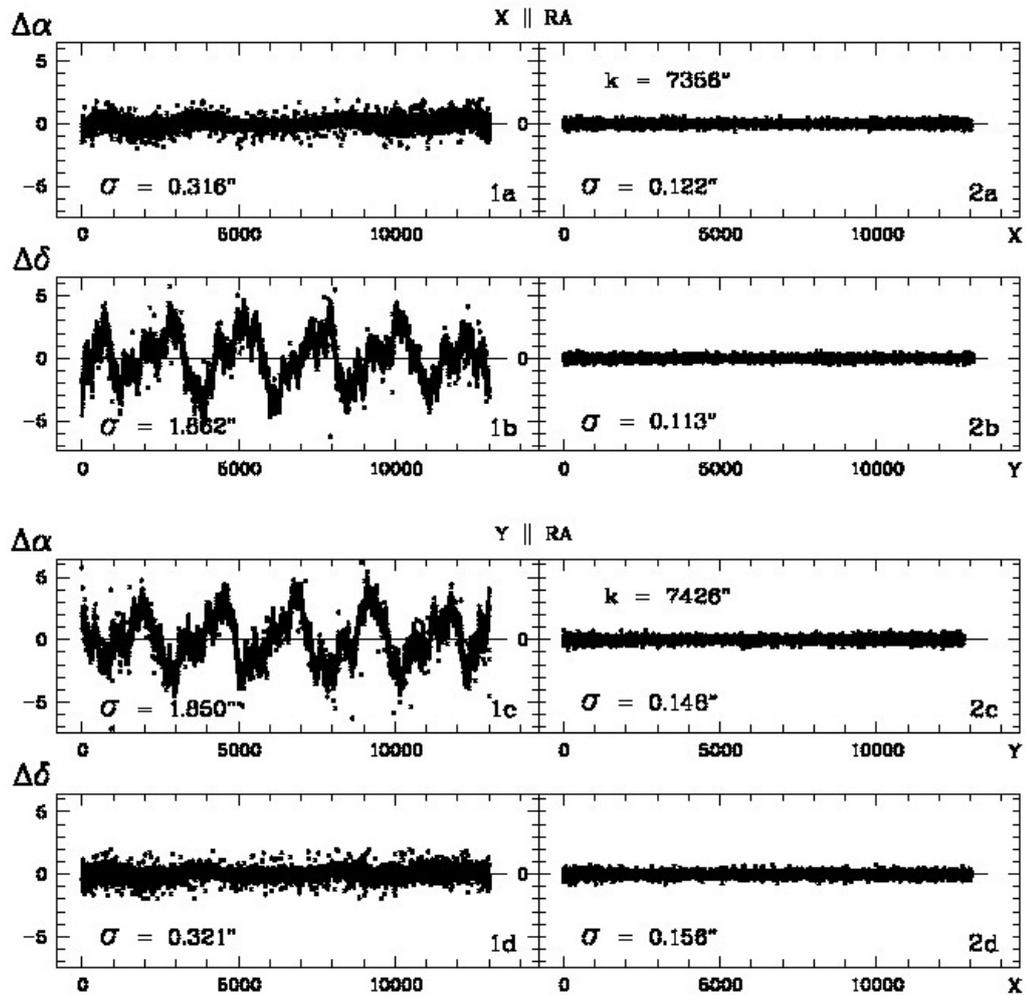

**Figure 1.** Results of the astrometric solution for test plate in two modes of scanning: normal (1a, 1b) and at 90° turn (1c, 1d); before the correction for scanner systematic errors (left) and after the correction (right).

For better statistical reliability of results scans of 23 plates with four-fold overlapping were processed separately for two modes of scanning and two test catalogues of 437,946 and 433,726 stars were obtained, covering the RA-zone from $18^h$ to $24^h$, and DEC-zone from 56° to 64°.

Values of errors $\sigma_\alpha, \sigma_\delta, \sigma_{Bph}$ of test catalogues and their trend depending on magnitude B are shown on Fig. 2. Two upper panels **a** and **b** correspond to two catalogues, the third panel corresponds to the catalogue, obtained by averaging the data of the first two catalogues and errors are obtained by averaging the modules of semi-differences of the same data for 391,852 common stars. Here, the values, corresponding to internal errors of the every catalogue are also given.

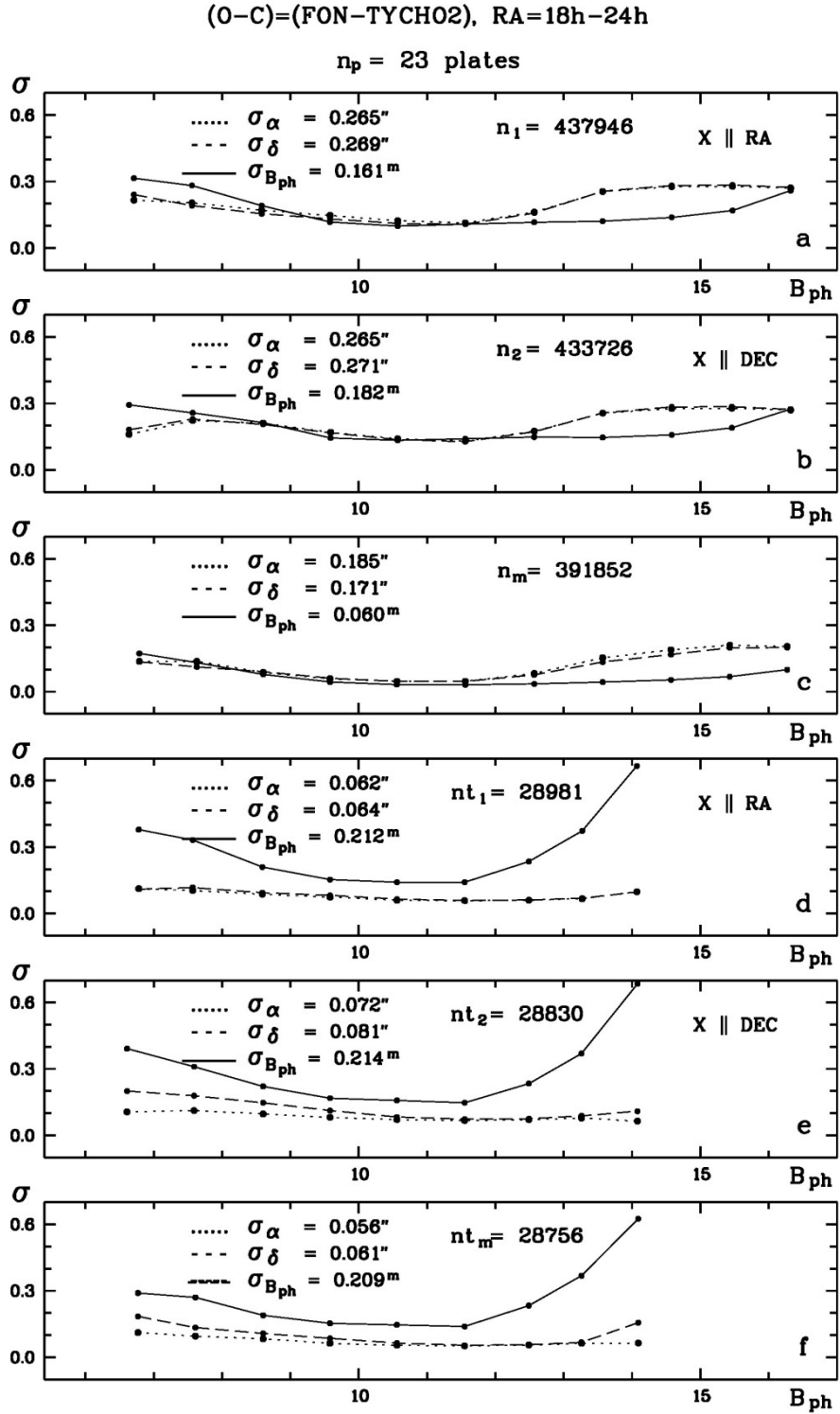

**Figure 2.** The trend of internal errors $\sigma_\alpha$, $\sigma_\delta$ and $\sigma_B$ vs the star magnitude $B_{ph}$ for 23 processed FON plates. Three upper panels show data for normal scanning (**a**), scanning at 90° turn (**b**) and averaged data for two scanning modes (**c**). Three lower panels show results of comparison for three above said cases with TYCHO-2.

For both modes of scanning the magnitudes of errors are approximately equal; for the averaged catalogue internal positional errors decrease approximately by 40-50% and the internal errors of star magnitudes decreased by two times. Three test catalogues were compared with UCAC4 catalogue, results are shown on panels **a,b,c** on the Fig. 3.

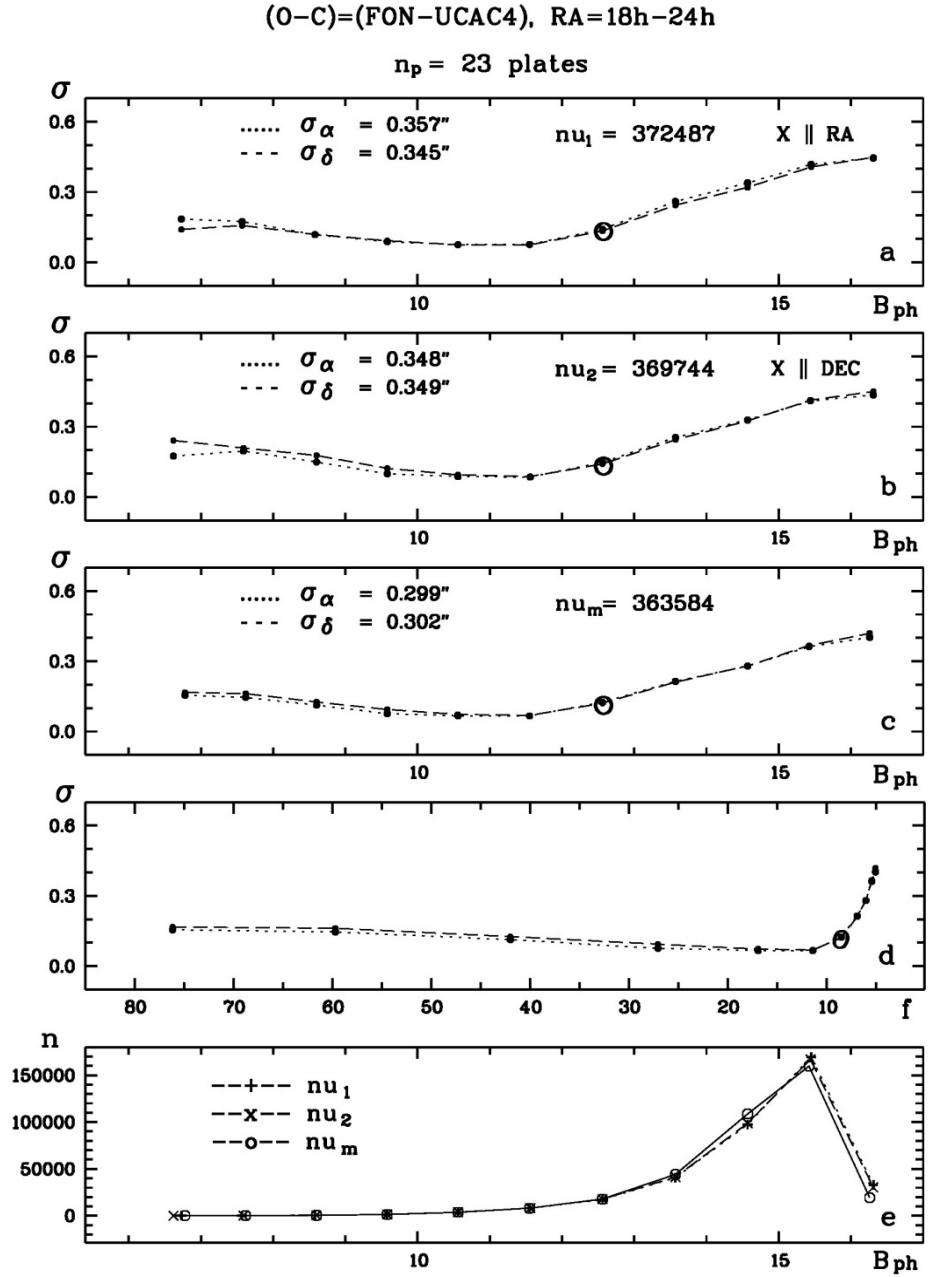

**Figure 3.** The trend of positional errors on external convergence, derived from comparison with UCAC4. On panels **a, b, c** the errors are given in relation to magnitudes B, on the **d**-panel in relation to diameters of star images $f_S$. The lower panel **e** shows the distribution of star amounts with magnitudes for three comparison cases shown on **a, b, c**-panels.

The external convergence was evaluated by comparison of three catalogues with UCAC4 [21]. Results of comparison in the form of the distribution of averaged positional errors along the scale of magnitudes $B_{ph}$ are given on on **a, b**-panels of Fig. 3. The correlation of positional errors $\sigma_\alpha$, $\sigma_\delta$ with the diameters of star images is shown on the **d**-panel. The distribution of stars common with UCAC4 in relation to star magnitudes is given on the **e**-panel.

Note, that the external positional errors for the averaged catalogue are less approximately by 10% than the same ones for separate test catalogues . Analysis of the distribution of positional errors (**d**, Fig.3) in comparison with the same one shown on Fig.2 gives the next results: the value of an error for faint stars strongly depends on diameters $f_S$ (FWHM) of their images. The significant increase of positional errors starts from 8 pixels and below, that corresponds to $B_{ph} \approx 12.5^m$. The last means that scanning the plates in position rotated by 90° leads to increasing the positional errors by 20% relative to the reference frame. Thus, the use of two positions improves catalogue upto 10% at 100% increasing of work and data size. Therefore the catalogue of 60-degree zone is obtained at normal scanning position of plates.

**Separation of stars into two exposition samples**. The FON plates were obtained with two expositions: the long and short of 16-20 minutes and 30-60 second respectively. For astrometric catalogue star images of short exposition are not used and should be excluded at the initial step before the astrometric solution [3]. Stages and functional dependences of different parameters in the separation of detected objects into two samples are shown on Fig.4. Upper panels demonstrate the correlation between instrumental photometric values of long and short expositions $m_2$ and $m_1$ at initial and final stages of separation. Differences of magnitudes $\Delta m$, reduced to the mean value, differences of distances between centers of images $\Delta r$, differences of rectangular coordinates $\Delta X$, $\Delta Y$ are given in relation to $m_1$ (panel **a**), the distance from the center of the plate R, rectangular coordinates Y, X (panels **b, c, d, e, f**). The lower panels (**g, h, j**) show real and pre-calculated histograms of distribution $\Delta m$, $\Delta X$, $\Delta Y$, solid and dashed lines, respectively. Note, that the value of the $\Delta X$, $\Delta Y$ differences rotation in relation to the center of rectangular coordinates Y, X depends on the declination of the plate. The mutual rotation of two frames is absent at the equator.

**Corrections for scanner systematic errors**. Both types of scanners have systematic errors affected the positions, especially large ones are along the Y-axis, which coincides with the CCD - line movement direction. The amplitude of differences between observed and catalogue coordinates reaches $\Delta_{\alpha\delta} = \pm 2.5"$ for Epson Expression 10000 and $\Delta_{\alpha\delta} = \pm 5"$ for Microtek ScanMaker 9800XL TMA. Here, the correction of X, Y for scanner systematic errors is made next way. The length of the plate along the Y-axis $L_Y \approx 13,000$ px is divided by the number of reference stars $N \approx 7,300$, which gives us the initial step of approximation: $s = L_Y/N \approx 2px$. This means that every step **s** must have at least one TYCHO-2 star. If step **s** has two or more reference stars reference points are calculated for the middle of the step as a mean deviation $\Delta X = \Delta\alpha/M$ and $\Delta Y = \Delta\delta/M$ where M is the scale of scanning ( $1px \approx 2.17"$ ) from the true position on the plate; if the reference stars in the step are absent deviations of reference points $\Delta X$, $\Delta Y$ are obtained by interpolation of two adjacent steps. For determined stars $\Delta X$, $\Delta Y$ are calculated by interpolation of adjacent reference points.

Note that the complete reduction of rectangular coordinates X,Y into the reference system is done in several consecutive approximations, in each cycle the step is increased by the value of the step itself. The approximations converge in the fourth step.

**Accounting the magnitude equation.** In calculating the tangential coordinates the special attention is given to the accounting of the magnitude equation mdtX and mdtY. It was found that for the plates exposed on astrographs the magnitude equation becomes significant for stars from $B \approx 11^m$ and its affection increases with the brightness of stars.

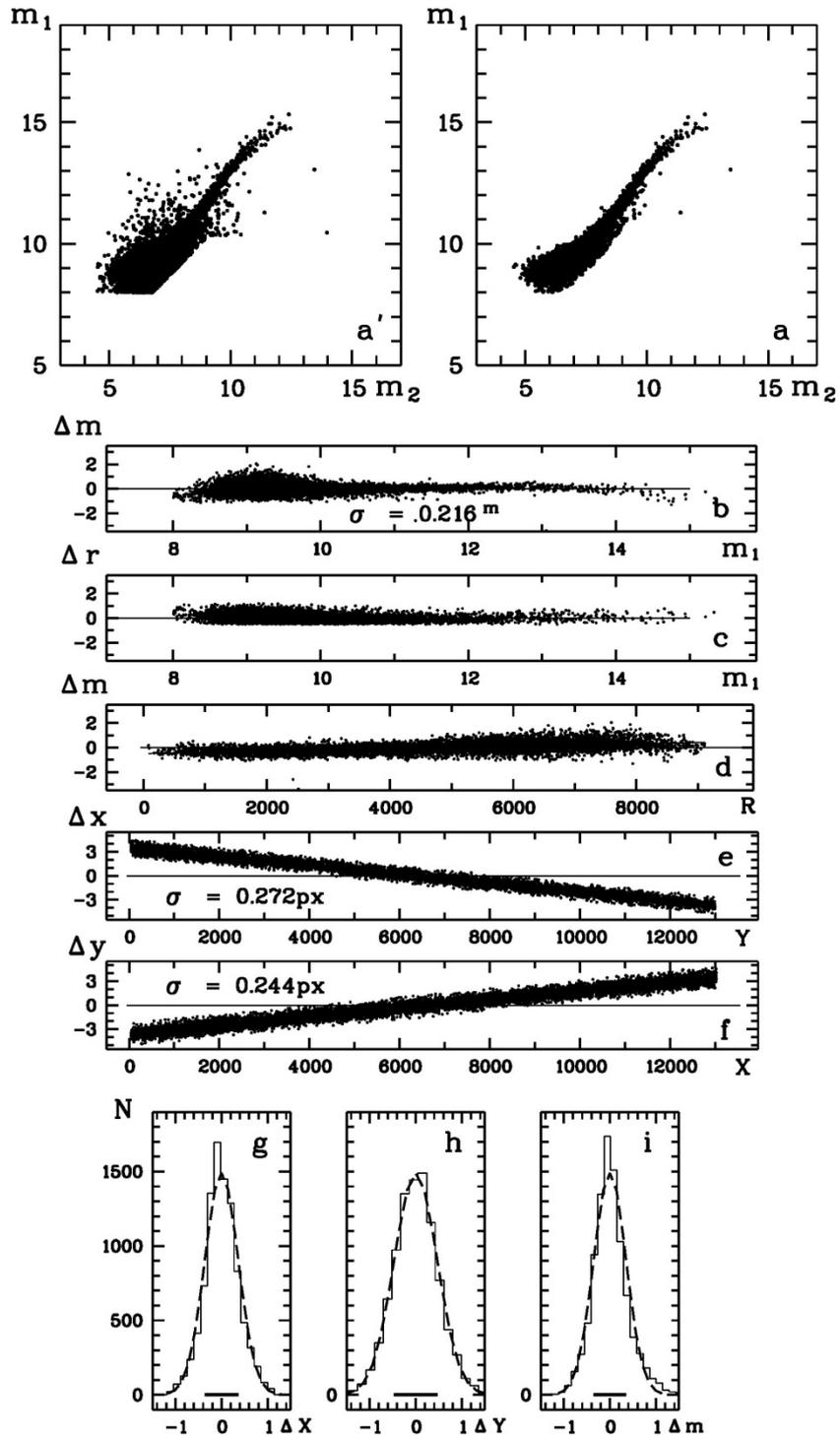

**Figure 4.** Stages of processing and functional dependences of different parameters in separation of star images into two exposition samples on the example of the test plate.

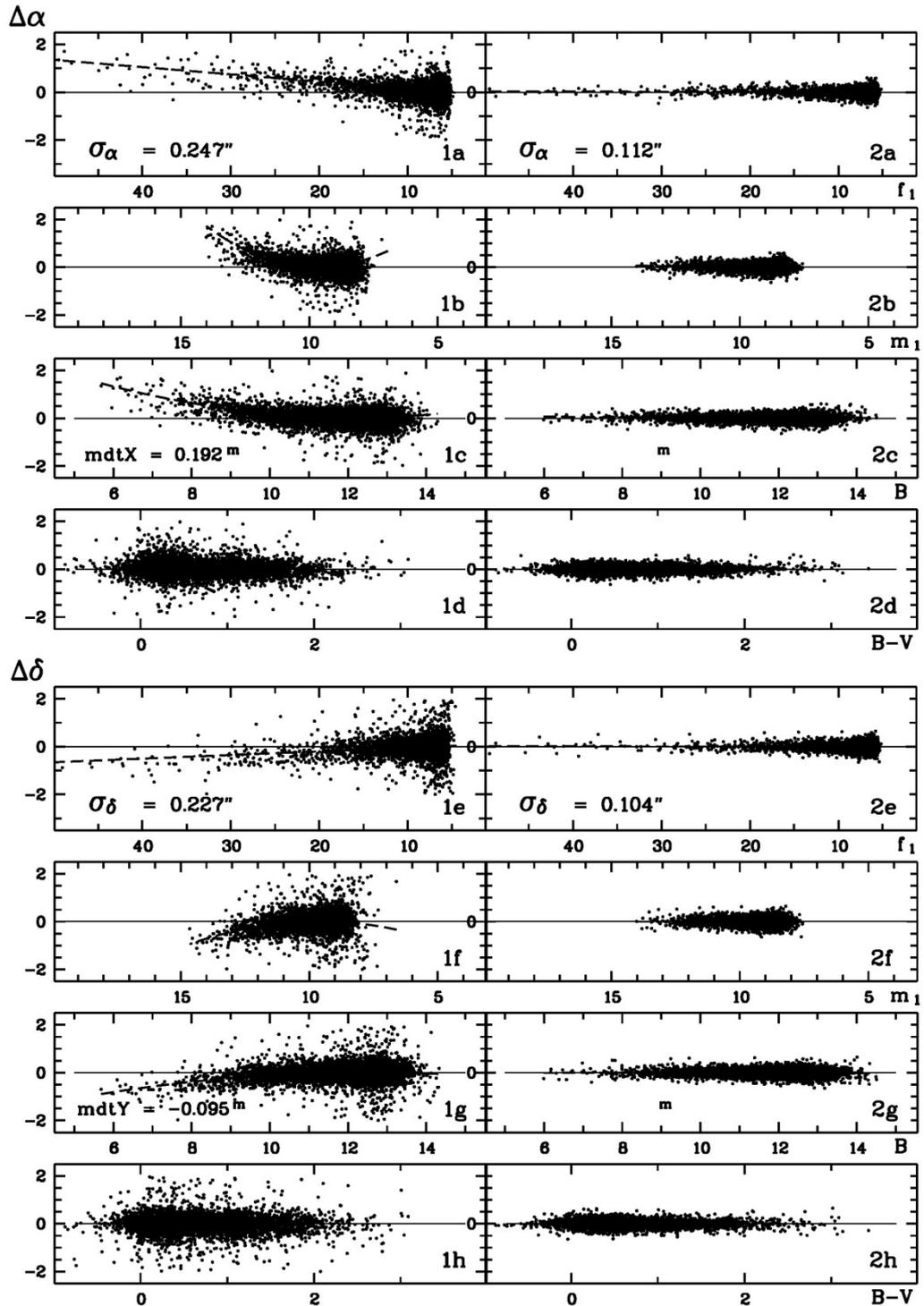

**Figure 5.** On the left side the diagnostics of the magnitude equation is shown in relation to star image diameters $f_1$ (panels **a, e**), instrumental photometric measurements $m_1$ (panels **b, f**), star magnitudes B (panels **c, g**) and color indices B-V (panels **d, h**). On the right side the results of the final astrometric solution are given.

Fig. 5 shows the magnitude equation for long exposition of the test plate. Left panels demonstrate differences $\Delta\alpha$, $\Delta\delta$ between observed and catalogue positions in relation to star image diameters $f_{\frac{1}{2}}$, photometric measurements $m_1$, magnitudes B and color indices B-V of TYCHO-2 before the corrections for systematic errors of the scanner. From plots it is obvious that the magnitude equation is linear on all sections of $f_S$ and B and there is quadratic dependence on photometric measurements $m_1$. Panels on the right side of the Fig.5 show the trend of residual differences after the correction for the magnitude equation and the errors of the scanner.

**Astrometric solution algorithm.** For the fields of plates with dimensions 8°x8° tangential coordinates are calculated by equations (1) in preliminary stages and in final solution:

$$\xi_i = a_1 + a_2 X_i f_i + a_3 Y_i f_i + a_4 R_i m_i + a_5 f_i + \sum b_{lm} X_i^l Y_i^m, \quad (k=0 \div 6, m=0 \div 6, k+m=n, n=1 \div 6)$$
$$\eta_i = c_1 + c_2 X_i f_i + c_3 Y_i f_i + c_4 R_i m_i + c_5 f_i + \sum d_{lm} X_i^l Y_i^m, \quad (k=0 \div 6, m=0 \div 6, k+m=n, n=1 \div 6)$$
(1)

Here, i = 1,2,…N – number of reference stars; $X_i$, $Y_i$ and $R_i$ – rectangular coordinates and distances of stars from the centers of plates; $m_i$ – photometric measured data of stars; $f_{\frac{1}{2}i}$ – diameters of star images; coefficients $a_2$, $a_3$, $a_4$ and $c_2$, $c_3$, $c_4$ define coma affects, coefficients $a_5$, $c_5$ – taking into account the magnitude equation, which is calculated separately; coefficients of the full sixth-order polynomial $b_{lm}$ and $d_{lm}$ (27 terms) in the generalized case describe the aberrations of telescope optics together with the additional systematic errors of the scanner.

Fig.6 shows the results of the test plate processing. On the left side the trend of telescope systematic errors $\sigma_\alpha$, $\sigma_\delta$ over the plate field is shown. Right panels demonstrate the trend of the residual differences $\Delta\alpha$, $\Delta\delta$. Negative and positive values of differences are shown by horizontal and vertical strokes which have linear dimensions according to the scale of values presented on the figure. Errors are obtained by averaging within 250x250 px cells.

**Photometry with two expositions.** The combination of two characteristic curves of both expositions gives the possibility to spread the characteristic curve onto the whole interval of magnitudes for the photoemulsion of the plate. Photoelectric $B_{pe}$ magnitudes from catalogues [12,13] were used as Johnson photoelectric standards. Figures 7 and 8 show the algorithm of building the characteristic curve of the astronegative accounting the photometric field error and using the data of two expositions on the example of one plate. Left panels of Fig. 7 up to down show the relation between two expositions for star image diameters and instrumental photometric data. $f_2$ is the diameter and $m_2$=35 sec is the exposure time for the short exposition, and $f_1$, $m_1$= 980 sec are the same data for the long one. Panels **a,c** concern all common stars of two expositions, the panel **e** shows data of stars used as standard.

The right upper panel **b** shows classic characteristic curves for long (1) and short (2) expositions build as diameters vs star magnitudes. The difference of two expositions derived in the scale of photoelectric standards $B_{pe}$ is $\Delta B = 3.181^m$. The errors of separately determined photometric evaluations for classic characteristic curves are $\sigma_1 = \pm 0.180^m$ and $\sigma_2 = \pm 0.260^m$ for long and short expositions respectively.

The panel **d** shows the same characteristic curves but built as the relations of instrumental photometric evaluations $m_1$, $m_2$ vs standards $B_{pe}$. The errors derived from the above said relations are $\sigma_1 = \pm 0.109^m$ and $\sigma_1 = \pm 0.159^m$.

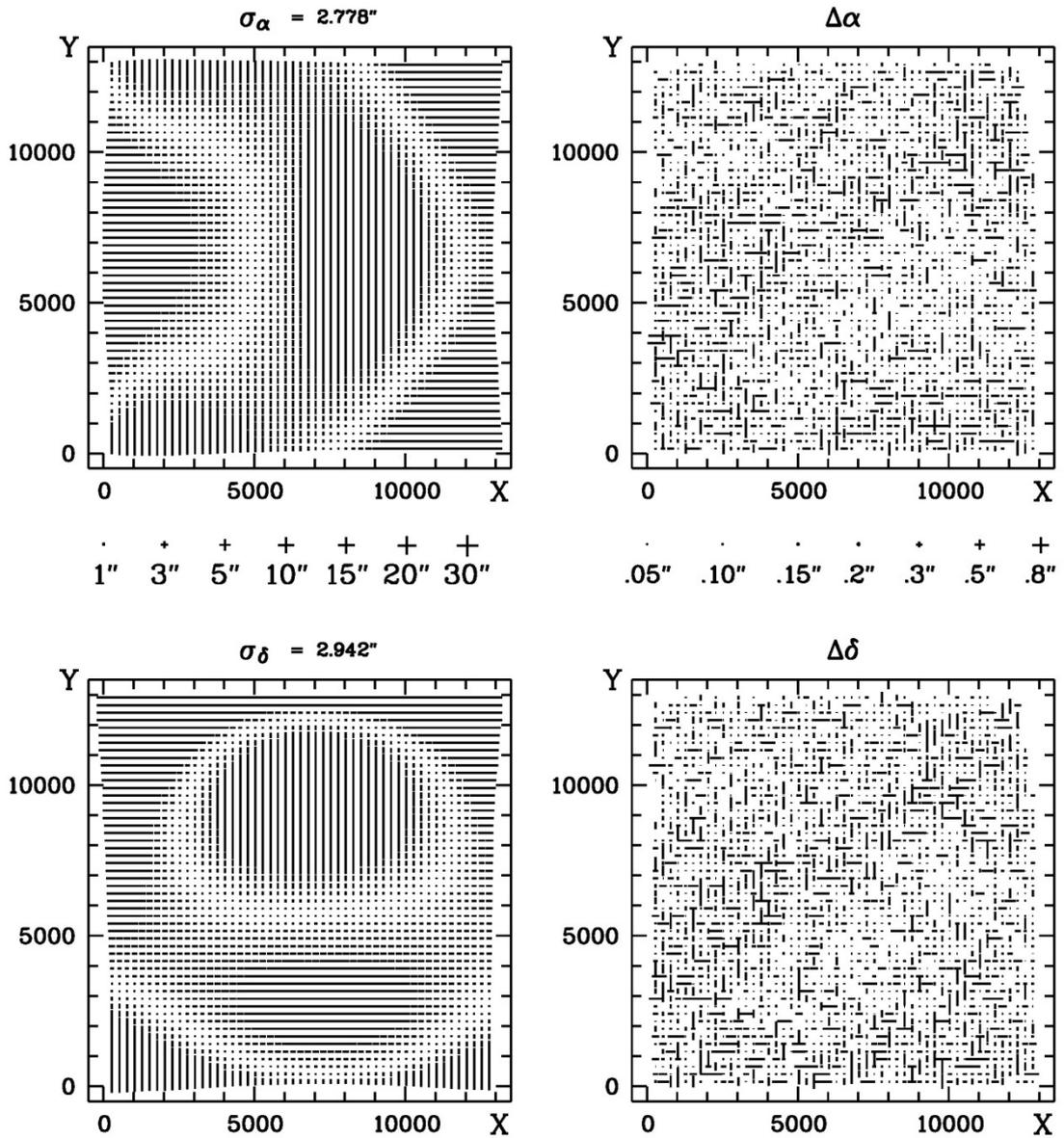

**Figure 6.** The results of astrometric solution for the test plate. Left side plots show the trend of telescope systematic errors $\sigma_\alpha$, $\sigma_\delta$ over the field of the plate. Right side plots demonstrate the trend of the residual differences $\Delta\alpha$, $\Delta\delta$. Negative and positive values of differences are shown by horizontal and vertical strokes which have linear dimensions according to the scale of values presented on the figure. Errors are obtained by averaging within 250x250 px cells.

The panel **f** gives the characteristic curve as the combination of two exposures. It was used for further reduction of photometric evaluations $m_1$ into $B_{pe}$. The accuracy of its building is $\sigma_1 = \pm 0.115^m$, the contrast ratio of the emulsion is $\gamma = -0.763$ (or $\approx 37°$).

The approximation of characteristic curves for all 120 plates was made by the rms solution of the equations (2):

$$B_i = e_1 + e_2 X_i + e_3 Y_i + e_4 R_i + e_5 R_i^2 + e_6 R_i^4 + \sum f_n m_i^n, \quad (n=1,2,\ldots 5). \qquad (2)$$

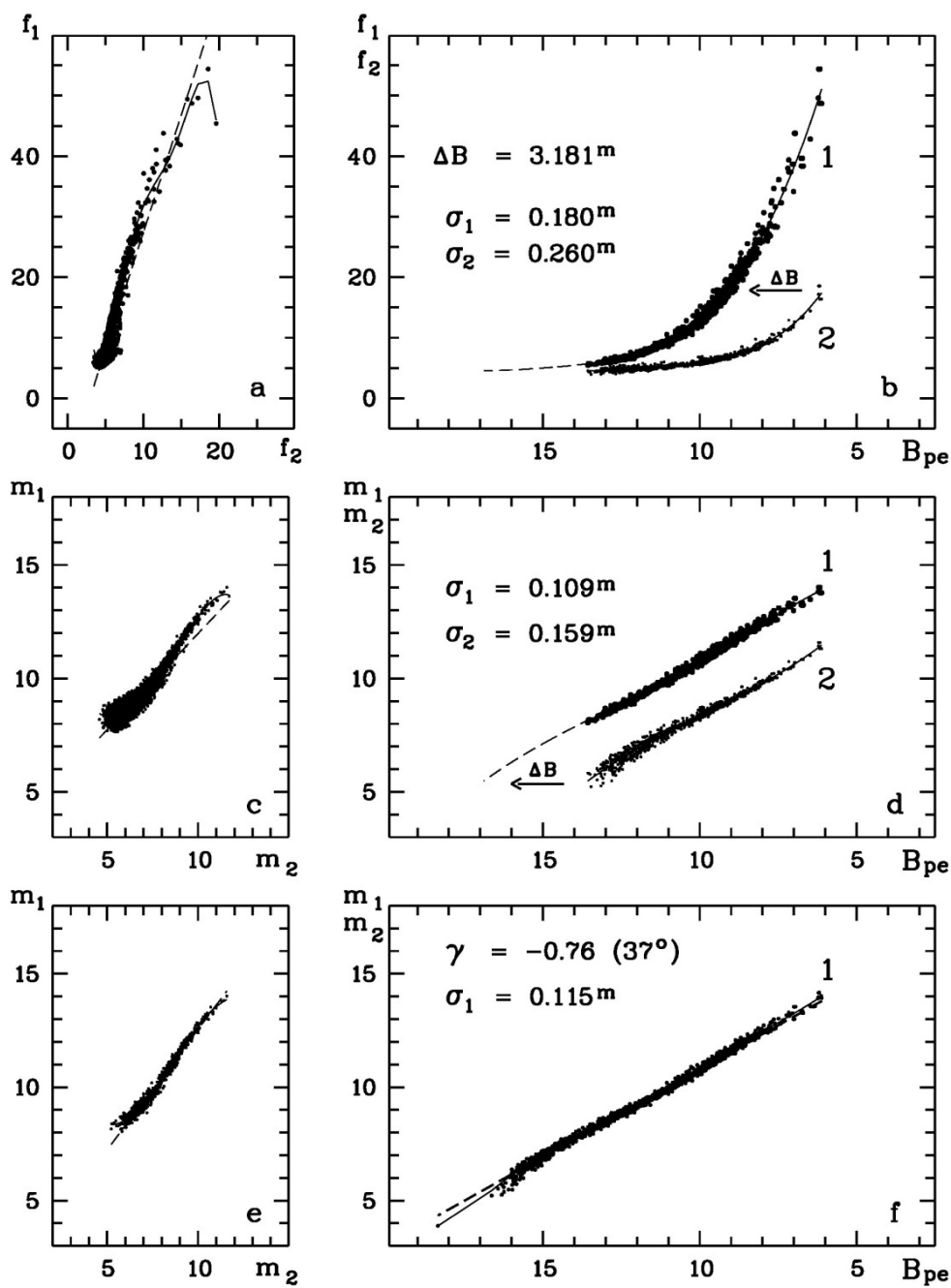

**Figure 7**. Star photometry with the combination of two exposures. Examples of relations are given for the test plate.

Here, i = 1,2,...N is the number of photoelectric data for standard star on the plate; $X_i$, $Y_i$ and $R_i$ are the rectangular coordinates and the distances from the center of the plate; $m_i$ – instrumental photometric evaluations; coefficients $e_2$, $e_3$, $e_4$, $e_5$, $e_6$ define the field photometric equation; $f_1$, $f_2$, $f_3$, $f_4$, $f_5$ describe the functional form of the characteristic curve. The equation (2) was chosen as minimizing the errors of the reduction in the best way.

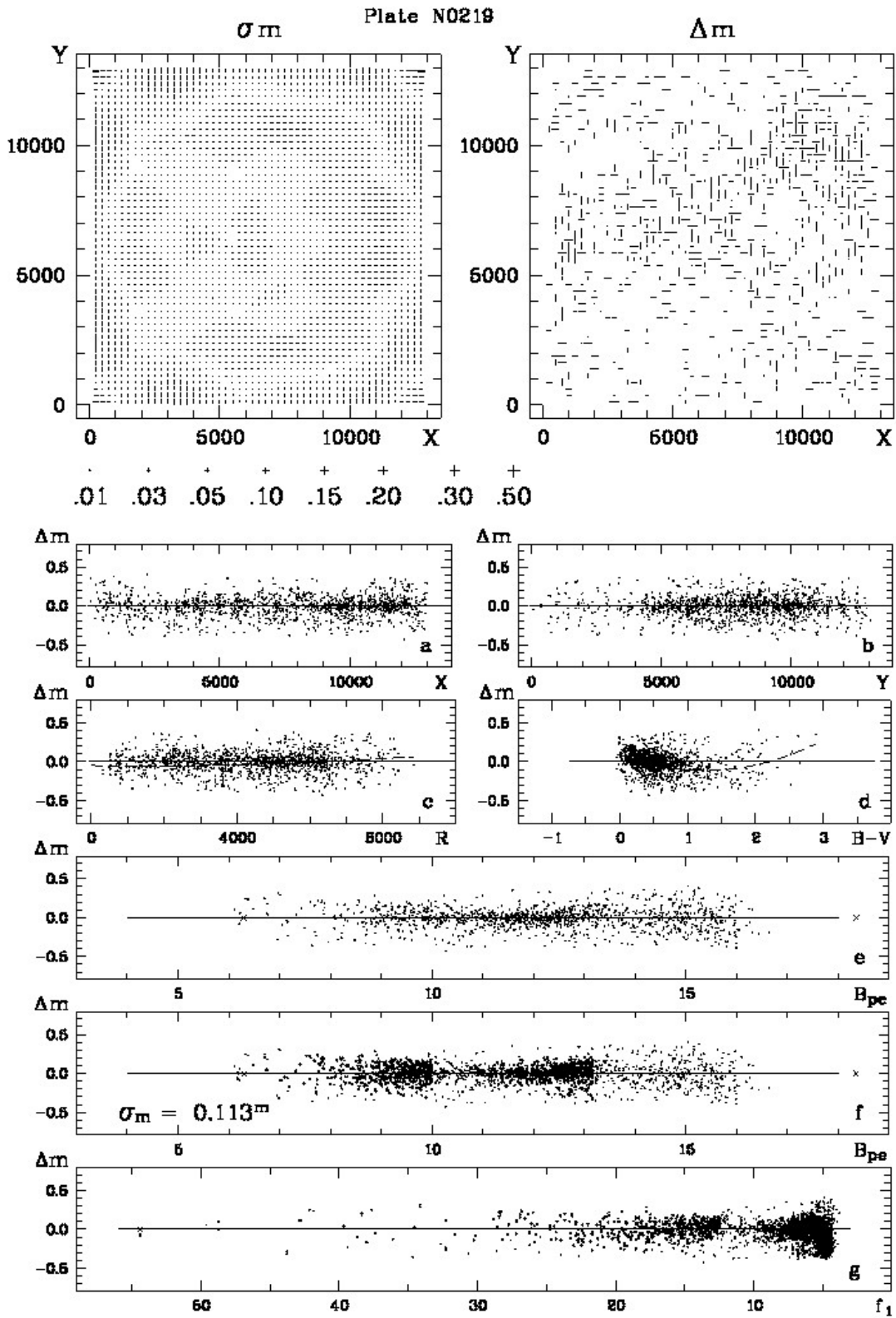

**Figure 8.** Test example. Field photometric equation (upper panels and **a, b, c, d** panels) and trends of differences between observed $B_{ph}$ and standard $B_{pe}$ magnitudes $\Delta m$ (**e, f, g** panels).

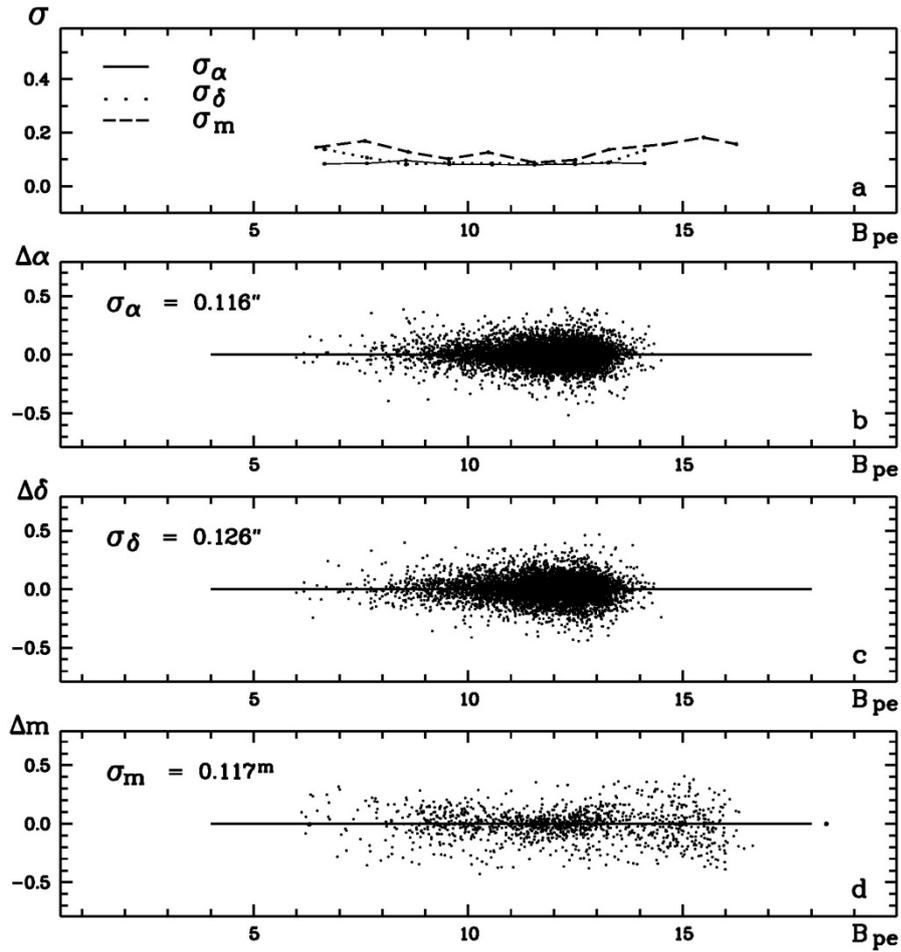

**Figure 9.** The trend of errors $\sigma_\alpha$, $\sigma_\delta$, $\sigma_m$ (**a**), differences $\Delta\alpha$, $\Delta\delta$ (**b, c**) and $\Delta m$ (**d**) in relation to star magnitudes.

The upper panels of Fig.8 show the distribution of photometric equation errors $\sigma_m$ (left) and individual differences $\Delta m$ obtained as $B_{pe} - B_{ph}$ (right) over the plate field. The same differences are also given in relation to coordinates X,Y (**a,c** panels), distances from the center R (**c**-panel) and color indices B-V (**d**-panel). The panel **e** shows the trend of $\Delta m$ for standard stars, **f** one shows the same differences plus stars of TYCHO-2 ($B_T \leq 11^m$, $0.0^m \leq (B-V)_T \leq 1.0^m$) additionally used for the determination of the field photometric errors. The lower panel **g** demonstrates the trend of $\Delta m$ in relation to diameters of first exposition star images $f_1 = f_{½}$.

The summarized results are shown on Fig.9. The trend of errors $\sigma_\alpha$, $\sigma_\delta$, $\sigma_m$ in relation to star magnitude is given on **a**-panel. The same dependences for differences between derived and standard coordinates $\Delta\alpha$, $\Delta\delta$, $\Delta m$ are shown on **b,c** and **d** panels respectively.

**Catalogue of positions and B-magnitudes in 60th declination zone of the FON project**

The first version of catalogue includes 1,108,603 stars and galaxies and was created in 2011 using combinations of scans obtained in two positions [9]. The second one (hereafter FON60v2) is created by processing of 120 normally scanned plates and includes 1,263,932 stars and galaxies down to $B \leq 16.5^m$ (photographic magnitudes in Johnson system) at the epoch

1984.76±0.50. The most of plates are 30x30 cm, scans have dimensions 13,000x13,000 px. The total amount of detected objects is 7.2 millions.

In the overlapping areas of the plates identification and selection of candidates for the stars and galaxies were made by the following criteria:
1. on coordinates: the differences should not exceed half-pixel (0.5px ≈ 1.1");
2. on magntudes: the differences should not exceed ±2$^m$ , taking into account variable stars.

The candidate was added to the catalogue if it is found on more than one plate.

Table 3 describes the distribution of internal positional and photometric errors ($\sigma_\alpha$, $\sigma_\delta$, $\sigma_{Bph}$), diameters of star images $f_{1/2}$, intencity values in the centers of object images cInt and number of objects n in dependence on magnitude intervals. The bottom row of the table shows the average values of the above variables.

Table 3. Distribution of internal positional and photometric errors ($\sigma_\alpha$, $\sigma_\delta$, $\sigma_{Bph}$), diameters of star images $f_{1/2}$, intencity values in the centers of object images cInt and number of objects n in dependence on magnitude intervals.

|    | $B_{ph}$ | $\sigma_\alpha$ | $\sigma_\delta$ | $\sigma_{Bph}$ | FWHM | cInt  | n         |
|----|----------|-----------------|-----------------|----------------|------|-------|-----------|
| 1  | 5.77     | 0.107           | 0.118           | 0.292          | 74.3 | 133.4 | 6         |
| 2  | 6.70     | 0.199           | 0.202           | 0.285          | 74.2 | 155.7 | 167       |
| 3  | 7.56     | 0.199           | 0.176           | 0.246          | 59.6 | 152.2 | 656       |
| 4  | 8.58     | 0.163           | 0.148           | 0.180          | 40.4 | 149.4 | 1,886     |
| 5  | 9.57     | 0.140           | 0.124           | 0.115          | 25.6 | 146.0 | 4,799     |
| 6  | 10.57    | 0.117           | 0.107           | 0.098          | 15.9 | 138.8 | 11,855    |
| 7  | 11.56    | 0.110           | 0.105           | 0.103          | 10.8 | 127.1 | 26,008    |
| 8  | 12.57    | 0.153           | 0.151           | 0.114          | 8.1  | 107.6 | 56,655    |
| 9  | 13.57    | 0.255           | 0.257           | 0.127          | 6.7  | 81.4  | 132,713   |
| 10 | 14.58    | 0.279           | 0.284           | 0.151          | 5.9  | 52.2  | 317,233   |
| 11 | 15.47    | 0.279           | 0.285           | 0.186          | 5.3  | 29.9  | 557,567   |
| 12 | 16.32    | 0.271           | 0.274           | 0.256          | 4.9  | 21.5  | 151,879   |
| 13 | 17.11    | 0.262           | 0.258           | 0.194          | 4.4  | 16.0  | 2,508     |
|    | 14.86    | 0.264           | 0.268           | 0.173          | 6.1  | 47.1  | 1,263,932 |

Fig.10 shows the trend of internal errors of astrometric solution on right ascension (dotted line), declination (dashed line) and photometric data (solid line) in dependence on magnitude intervals (**a**) and diameter intervals $f_S$ (**b**).

The comparison of FON60v2 with reference TYCHO-2 for 93,925 stars is shown on **c-panel**. The distribution of FON60v2-TYCHO-2 positional errors $\sigma_\alpha$, $\sigma_\delta$ , photometric values $\sigma_{BT}$, $\sigma_{BJ}$, diameters of star images $f_{1/2}$, intencity values in the centers of object images cInt and number of objects **n** in dependence on magnitude intervals is given in Table 4. It follows from the table that errors of astrometric solution for TYCHO-2 reference stars are better than $\sigma_{\alpha\delta} = \pm 0.060"$.

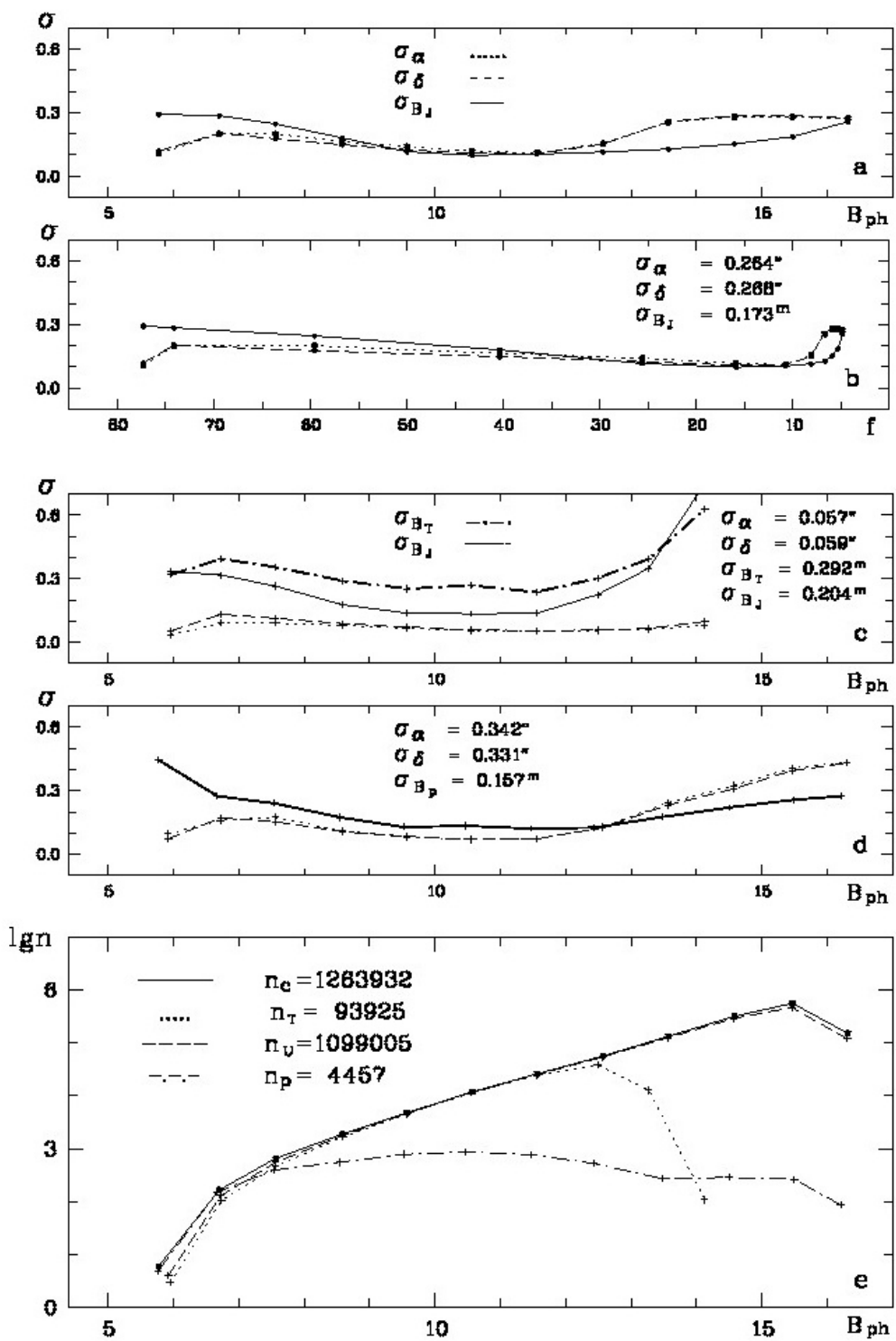

**Figure 10.** Errors of FON60v2 catalogue (1,263,932 stars and galaxies with $B \leq 16.5^m$): **a** and **b** - the trend of internal errors of astrometric solution on right ascension (dotted line), declination (dashed line) and photometric data (solid line) in dependence on magnitude intervals and diameter intervals $f_{1/2}$; **c** – the distribution of errors derived from the comparison FON60v2-TYCHO-2; **d** – errors $\sigma_\alpha$, $\sigma_\delta$ derived from comparison FON60v2-UCAC-4 and photometric

errors $\sigma_B$ derived from comparison with photoelectric standards; **e** – the distribution of comparison stars with magnitudes.

Table 4. The distribution of FON60v2-TYCHO-2 errors in dependence on magnitude intervals

|    | $B_{ph}$ | $\sigma_\alpha$ | $\sigma_\delta$ | $\sigma_{BT}$ | $\sigma_{BJ}$ | FWHM | cInt | n |
|----|------|-------|-------|-------|-------|-------|--------|--------|
| 1  | 5.95 | 0.035 | 0.053 | 0.319 | 0.334 | 65.38 | 129.48 | 3 |
| 2  | 6.73 | 0.090 | 0.132 | 0.394 | 0.318 | 72.22 | 154.97 | 104 |
| 3  | 7.56 | 0.090 | 0.113 | 0.355 | 0.265 | 58.94 | 152.67 | 468 |
| 4  | 8.59 | 0.080 | 0.088 | 0.290 | 0.177 | 39.80 | 150.35 | 1,671 |
| 5  | 9.57 | 0.065 | 0.072 | 0.253 | 0.138 | 25.57 | 146.22 | 4,581 |
| 6  | 10.56 | 0.055 | 0.058 | 0.268 | 0.134 | 15.93 | 138.87 | 11,444 |
| 7  | 11.55 | 0.053 | 0.055 | 0.237 | 0.136 | 10.78 | 127.39 | 24,476 |
| 8  | 12.49 | 0.056 | 0.057 | 0.303 | 0.225 | 8.25 | 109.25 | 38,246 |
| 9  | 13.27 | 0.062 | 0.065 | 0.393 | 0.349 | 7.02 | 89.76 | 12,823 |
| 10 | 14.13 | 0.079 | 0.098 | 0.630 | 0.750 | 6.20 | 66.69 | 109 |
|    | 11.88 | 0.057 | 0.059 | 0.292 | 0.204 | 11.41 | 117.68 | 93,925 |

The comparison of FON60v2 with UCAC4 is shown on the panel **d** Fig.10. Here, right acsension data ($\sigma_\alpha$) are marked with dotted line and declination ones ($\sigma_\delta$) with dashed line. Table 5 shows the distribution of FON60v2-UCAC4 positional errors $\sigma_\alpha$, $\sigma_\delta$, star image diameters $f_{½}$, intencity values in the centers of object images cInt and number of objects n in dependence on magnitude intervals. The positional errors FON60v2-UCAC4 are $\sigma_{\alpha\delta} = \pm 0.34"$. They were determined by 1,099,005 cross-identified stars and galaxies.

Table 5. The distribution of FON60v2-UCAC4 errors in dependence on magnitude intervals.

|    | $B_{ph}$ | $\sigma\alpha$ | $\sigma\delta$ | FWHM | cInt | n |
|----|-------|-------|-------|-------|--------|---------|
| 1  | 5.92  | 0.097 | 0.073 | 66.75 | 132.72 | 4 |
| 2  | 6.72  | 0.160 | 0.171 | 72.53 | 156.58 | 130 |
| 3  | 7.56  | 0.174 | 0.153 | 58.73 | 153.13 | 552 |
| 4  | 8.58  | 0.109 | 0.106 | 40.04 | 150.00 | 1,773 |
| 5  | 9.57  | 0.081 | 0.084 | 25.65 | 146.09 | 4,672 |
| 6  | 10.56 | 0.069 | 0.069 | 15.94 | 138.88 | 11,587 |
| 7  | 11.56 | 0.072 | 0.072 | 10.78 | 127.34 | 25,232 |
| 8  | 12.56 | 0.129 | 0.124 | 8.13 | 108.26 | 54,049 |
| 9  | 13.56 | 0.242 | 0.232 | 6.69 | 82.18 | 124,550 |
| 10 | 14.57 | 0.324 | 0.308 | 5.91 | 53.05 | 290,056 |
| 11 | 15.46 | 0.406 | 0.395 | 5.34 | 30.73 | 465,642 |
| 12 | 16.31 | 0.434 | 0.430 | 4.87 | 21.77 | 118,904 |
| 13 | 17.10 | 0.434 | 0.453 | 4.38 | 16.16 | 1,854 |
|    | 14.78 | 0.342 | 0.331 | 6.14 | 49.39 | 1,099,005 |

Photometric accuracy of resulted catalogue was determined by comparison of derived star magnitudes of FON60v2 with photoelectric magnitudes of 4,457 stars from reference catalogues. The results are given on the panel **d** Fig. 10 (solid line).

FON60v2 is located on the web-page of the MAO NASU web-site ftp://ftp.mao.kiev.ua/pub/astro/fon-2.0-zone-60/. The catalogue includes J2000 positions α, δ and B-magnitudes of 1,263,932 stars and galaxies down to B ≤ 16.5$^m$ at the epoch 1984.76±0.50. The catalogue contains also the positional errors, number of definitions, additional data of averaged star image diameters $f_{½}$ and intensity values in the centers of images cInt for every object.

## Conclusions

The thorough preliminary studies of scanned stuff processing features give a possibility to find the best algorithm of astrometric and photometric solutions in digitizing the photographic archives with widely spread models of commercial scanners. The application of their results and developed software make it possible to create extensive star catalogues without any practical losses in positional accuracy. The comparison of different models of scanners shows, that investigation of scanner behavior and the development of scanner errors model are important in the projects involving positional determinations. Properly selected processing algorithm not only can preserve the accuracy but significantly reduces the complexity and time-consuming of the project.

The second version of 60th declination zone catalogue FON60v2 was prepared based on the results of preliminary studies. It contains J2000 positions α, δ and B-magnitudes of 1,263,932 stars and galaxies down to B ≤ 16.5$^m$ at the epoch 1984.76±0.50. The positions of the objects are obtained in the TYCHO-2 reference system. The magnitudes are in the system of photoelectric standards. The internal accuracy of the catalogue in total is $\sigma_{\alpha\delta}$ = ±0.26" and $\sigma_B$ = ±0.17$^m$. For bright stars in the magnitude interval B = 8$^m$ -13$^m$ the errors are $\sigma_{\alpha\delta}$ = ±0.13" and $\sigma_B$ = ±0.12$^m$. The convergence of positions with TYCHO2 reference frame is $\sigma_{\alpha\delta}$= ±0.06" (by 93 925 stars), the convergence of B-magnitudes with photoelectric standards $\sigma_B$ = ±0.16$^m$ (by 4 458 stars). The comparison between FON60v2 and UCAC4 gives the evaluations $\sigma_{\alpha\delta}$ = ±0.34" (by 1,099,005 cross-identified stars and galaxies).

Algorithms and methods of digitized image processing together with the software specially developed in the Department of Astrometry MAO NAS of Ukraine are now successfully applied not only for catalogue creation from homogeneous series of photographic plates, obtained in reviews of sky, but for the processing of any plate with any observational parameters. It solves the problem of drawing the UkrVO photographic archives back into the field of scientific investigations.

**Acknowledgments.** The authors are grateful to the MAO NASU ACISS for the technical assistence. The authors thank Sc.D. A.I.Yatsenko for kindly providing scanned images of 60-zone plates.